\documentclass[acmsmall,screen,nonacm]{acmart}

\setcopyright{none}
\renewcommand\footnotetextcopyrightpermission[1]{}

\usepackage[utf8]{inputenc}
\usepackage[most]{tcolorbox}
\usepackage{xspace}
\usepackage{algorithm}
\usepackage{algpseudocode}
\usepackage{tikz}
\usetikzlibrary{arrows.meta,positioning}

\newcommand{\tool}{Dittany\xspace}

\begin{document}

\title{\tool: Contribution-Guided Repair Framework for MAS Applications}

\author{Hanqi Sun}
\email{51285902055@stu.ecnu.edu.cn}
\affiliation{
  \institution{East China Normal University}
  \city{Shanghai}
  \country{China}
}

\author{Yuchen Shao}
\email{ycshao@stu.ecnu.edu.cn}
\affiliation{
  \institution{East China Normal University}
  \city{Shanghai}
  \country{China}
}
\affiliation{
  \institution{Shanghai Innovation Institute}
  \city{Shanghai}
  \country{China}
}

\author{Hanyi Si}
\email{10235101481@stu.ecnu.edu.cn}
\affiliation{
  \institution{East China Normal University}
  \city{Shanghai}
  \country{China}
}

\author{Beibei Xu}
\email{51285900012@stu.ecnu.edu.cn}
\affiliation{
  \institution{East China Normal University}
  \city{Shanghai}
  \country{China}
}

\author{Lisheng Wang}
\email{52305902021@stu.ecnu.edu.cn}
\affiliation{
  \institution{East China Normal University}
  \city{Shanghai}
  \country{China}
}

\author{Chengcheng Wan}
\authornote{Corresponding author.}
\email{ccwan@sei.ecnu.edu.cn}
\affiliation{
  \institution{East China Normal University}
  \city{Shanghai}
  \country{China}
}
\affiliation{
  \institution{Shanghai Innovation Institute}
  \city{Shanghai}
  \country{China}
}

\renewcommand{\shortauthors}{Hanqi Sun et al.}

\begin{abstract}
LLM-based multi-agent system (MAS) applications can produce useful intermediate results while leaving task requirements unsatisfied. Repairing such executions requires identifying the work still needed and the completed results that can support it. We present \tool, a contribution-guided runtime repair framework for MAS applications. \tool derives expected contributions and acceptance conditions from task requirements and method specifications, and connects their dependencies in a contribution graph. Runtime checks locate unmet conditions and establish which results remain usable. These results guide action-level correction when available inputs suffice, or graph-level recovery when prerequisites or unfinished work need revision. Both repair scopes retain validated contributions and check repaired outputs before subsequent work uses them. We evaluate \tool on 30 applications across six MAS frameworks and 9,000 tasks. It detects runtime failures with 88.28\% precision and 79.18\% recall, and achieves 84.76\% task accuracy, improving over Native execution by 26.74 percentage points and the strongest of four recovery baselines by 21.28 points. Estimated deployment token consumption is 17.92\% above Native, but 24.19\% below Reflexion and 13.05\% below Iterative Rollback. Ablation results show that combining the two repair scopes achieves higher accuracy and lower token consumption than graph-level repair alone.

\end{abstract}

\maketitle
\pagestyle{plain}

\section{Introduction}
\subsection{Motivation}

Large language model (LLM)-based multi-agent systems (MASs) coordinate agents with different roles to complete complex tasks~\cite{wu2023autogen,qian2023chatdev,fourney2024magenticone}. During a run, agents delegate subtasks, invoke tools, and use the returned results to decide what to do next. We use action to refer to a single execution of a method, whether through an agent invocation or a tool invocation. MAS applications therefore depend on both the decisions agents make and the content they produce. These decisions and outputs can conflict with the application's requirements: an action may receive unsuitable inputs, an agent may be assigned a subtask beyond its capabilities, or a result may lack the evidence needed by subsequent actions. Such problems can leave task requirements unsatisfied even when the MAS application continues executing normally.

These failures affect both the reliability and the cost of MAS applications. They may cause the application to crash or allow the run to finish with an answer that does not satisfy the user request. Once an incorrect or incomplete result enters the shared context, later agents may use it in their own work. Meanwhile, other parts of the task may already have produced useful results. Retrying the latest action with the same inputs may leave missing information or unresolved constraints unaddressed, whereas restarting the entire run repeats work that was already correct and may reproduce the same failure. Repair therefore requires identifying which existing results remain valid and planning the additional or revised actions needed to satisfy the unmet task requirements.

\begin{figure}[H]
    \centering
    \includegraphics[width=\linewidth]{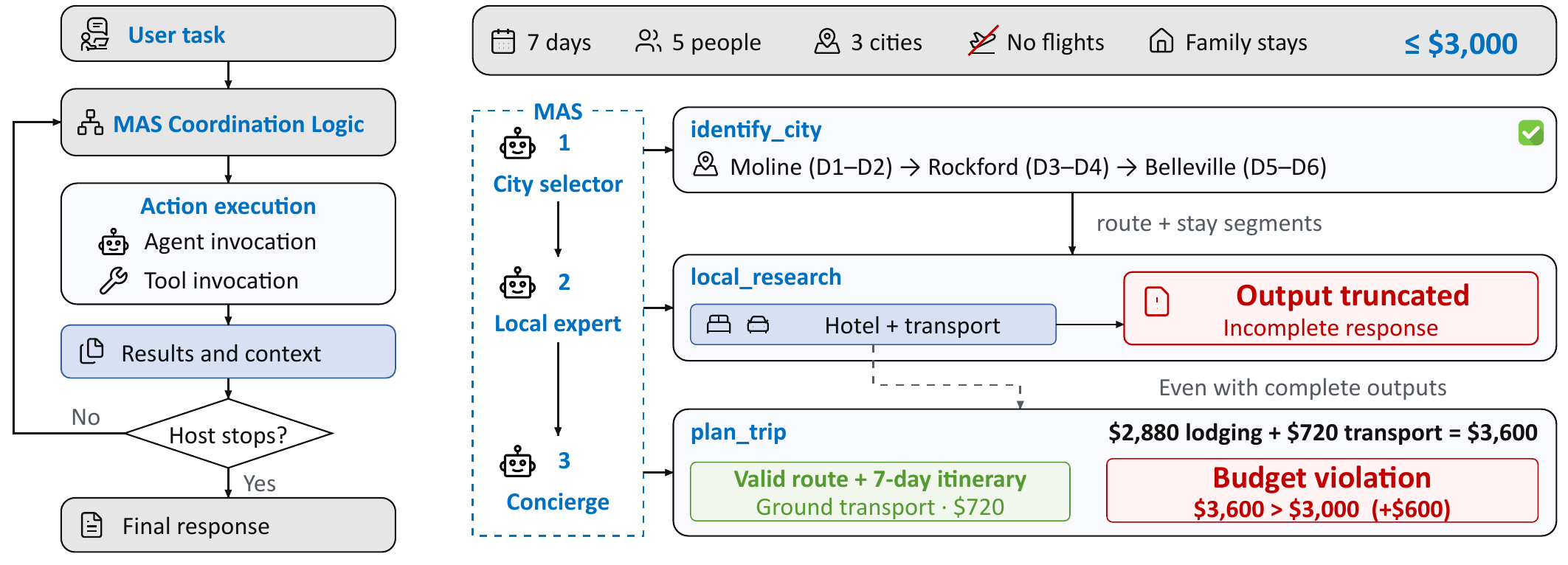}
    \caption{MAS runtime workflow (left) and failures in a travel-planning execution (right).}
    \label{fig:motivating-example}
\end{figure}

Consider the travel-planning execution in Figure~\ref{fig:motivating-example} (right). The user requests a seven-day trip for five people across three cities, with ground transport, family accommodation, and a budget of \$3,000. A city selector proposes Moline, Rockford, and Belleville; a local expert collects accommodation and transport options; and a concierge assembles the itinerary. The local expert first returns a truncated structured response, leaving the concierge without the complete accommodation and transport options needed to assemble the itinerary. Retrying this action can restore the structured response and allow itinerary assembly to proceed~\cite{cemri2025why}. During itinerary assembly, the concierge combines \$2,880 in lodging with \$720 in transport, exceeding the budget by \$600.

One repair option is to retain the selected transport and search for accommodation within the remaining \$2,280. Because none of the existing accommodation options satisfies both this budget and the family-accommodation requirements, this repair requires additional accommodation research.
Recovering this execution requires correcting the incomplete response and supplying new accommodation choices while retaining the usable route and transport results.

Existing techniques improve important aspects of agent reliability in MAS applications.  Reflection-based methods use execution feedback to generate critiques and adjust subsequent behavior~\cite{shinn2023reflexion}. Runtime supervision introduces an additional agent to monitor interactions and provide corrective guidance~\cite{lin2025supervisoragent}. Rollback-based repair restores an earlier state and re-executes the remaining trajectory with revised instructions~\cite{ma2025dover,lu2026mars}. Post-hoc diagnosis identifies failure-critical portions of an execution and generates guidance for subsequent recovery~\cite{zhao2026agenttether}. However, model-generated corrections can still leave application requirements unsatisfied, while rollback and repeated execution can redo useful work and introduce additional token costs. Applying a repair in a running MAS application also depends on the requirements of the application and the work already completed. We therefore investigate how to repair a partially failed execution of a MAS application: what work must be added or revised, which existing results remain usable, and how the application can continue from that state.

\subsection{Challenges}

\noindent\textbf{C1: How to detect failures beyond explicit execution errors?}
MAS failures can occur even when agents return well-formed outputs. A response may omit required information or make an unsupported claim, and individually valid results may violate a task requirement when combined. In the travel example, the accommodation and transport records are valid on their own, but their total cost exceeds the budget. Detecting such failures requires relating actions and results to method requirements and the user request. Some requirements can only be checked once results from several actions become available.

\noindent\textbf{C2: How to determine the scope of a repair?}
A failure may be detected in one action even though resolving it requires changes elsewhere. Retrying the action may suffice when its inputs are adequate, but missing or unsuitable inputs may require additional work. At the same time, other actions may already have produced results that remain useful. The challenge is to identify which actions need to be repeated, added, or revised, and which results can support the repair.

\noindent\textbf{C3: How to produce a repair that satisfies the task requirements?}
Knowing which actions need to change does not determine how to correct them. Agents may repeat the same mistake after receiving feedback, and a revised result may resolve one violation while introducing another. In the travel example, a cheaper accommodation option may satisfy the budget but fail to accommodate the whole family. A successful repair must satisfy the requirements of the affected action together with the constraints imposed by the task and subsequent actions.

\subsection{Contribution}

In this paper, we present \tool, a runtime repair framework for MAS applications. It uses a contribution graph to connect task requirements to the results and actions needed to satisfy them. Runtime checks identify unmet requirements and reusable results, which guide the choice of repair.

To tackle Challenge~1, \tool derives verification policies from the MAS specification and the user request. These policies check individual actions and results, as well as requirements involving several results. The checks identify execution errors, unsupported results, and violations of task requirements. Requirements awaiting evidence remain pending until the necessary results become available.

To tackle Challenge~2, \tool uses the contribution graph to identify the inputs and results needed to satisfy unmet requirements. Action-level repair corrects an action using its existing inputs; graph-level repair adds missing prerequisites or revises unfinished work. Both reuse results that satisfy the acceptance conditions of the work they support.

To tackle Challenge~3, \tool constructs repair candidates using available methods, retained inputs, and the acceptance conditions of the required results. It checks method choices, input bindings, and dependencies before execution, then verifies the resulting outputs against the relevant action and task requirements. Failed checks guide further revisions, and a final check assesses whether the completed response satisfies the user request.

We evaluate \tool on 30 applications across six MAS frameworks and 9,000 application--task instances. \tool achieves 88.28\% detection precision and 79.18\% recall. It achieves 84.76\% task accuracy, compared with 58.01\% for Native execution, 63.48\% for Reflexion, 58.21\% for Iterative Rollback, 48.39\% for SupervisorAgent, and 50.41\% for AgentTether. The gains are 26.74 percentage points over Native and 21.28 points over the strongest aggregate recovery baseline. Across all 9,000 tasks, estimated deployment tokens increase from 317.963M for Native to 374.948M for \tool, an increase of 17.92\%.

\section{Background}
\subsection{Runtime workflow of MAS}
\label{sec:background-runtime-flow}

An LLM-based multi-agent system (MAS) coordinates agents and tools to complete a user task. Frameworks used to build such applications include AutoGen/AG2, CrewAI, LangGraph, Google ADK, HelloAgents, and the OpenAI Agents SDK. Agents invoke tools, delegate work, and exchange results through the host's execution and communication interfaces.

In Figure~\ref{fig:motivating-example} (left), the host selects or dispatches actions according to its orchestration policy. An action is a single execution of a method, through an agent invocation or a tool invocation. Each action receives inputs from the user request and available execution context, performs the corresponding operation, and returns a result. Returned results update the execution context and may supply subsequent actions or become part of the final response. Sequential workflows order dependent work, parallel workflows combine results from separate branches, and dynamic workflows choose further work from runtime feedback. In each case, the application depends both on the behavior of an individual action and on how its results are used by subsequent work. The host's termination condition determines when this control process stops.

\subsection{Runtime Failures in LLM-Based Multi-Agent Systems}
\label{sec:background-runtime-failures}

Runtime failures in a MAS may become visible within an individual execution or through the collaboration among multiple executions. An agent or tool may fail to provide the result required by subsequent steps or the final response because of problems in input preparation, execution, or returned content.

Failures across executions can be less apparent. An intermediate result may appear plausible but lack sufficient support, leading a downstream agent to produce a well-formed continuation based on an unverified premise. Several agents may also complete their assigned work while their combined outputs leave part of the user request unresolved. The host MAS may therefore terminate with results that remain insufficient or unreliable for the task.

The shared task context allows these failures to propagate: later agents may reuse unsuitable or unsupported results in prompts, tool inputs, or the final response. Meanwhile, other agents or parallel branches may have produced valid and relevant results. A failed MAS execution can therefore contain an affected dependency chain alongside unaffected work that remains useful.

\section{An Illustrative Example}
{\label{sec:illustrative-example}

The travel-planning example illustrates two kinds of repair: correcting an action with the inputs already available and adding work to obtain missing inputs. In both cases, repair should reuse valid results and check that the revised itinerary meets the user request.

\noindent\textbf{Checking individual and combined results.}
The user requests a seven-day trip for five people across three cities, with ground transport, family accommodation, and a total budget of \$3,000. City selection produces a route, local research finds accommodation and transport options, and itinerary assembly combines them. Each result must meet its own requirements, such as the family-accommodation requirement. The completed itinerary must also meet requirements that involve several results, such as the total budget. Checking each accommodation and transport record separately cannot establish whether their combined cost fits the budget.

\noindent\textbf{Repairing an action with available inputs.}
Suppose city selection produces a valid route, but local research returns a truncated response. Itinerary assembly cannot proceed because some accommodation and transport information is missing. The inputs to local research are already available, so retrying this action can recover the missing information without repeating city selection. Once the new response passes its checks, itinerary assembly can continue.

\noindent\textbf{Adding work to obtain missing inputs.}
Now suppose itinerary assembly selects accommodation costing \$2,880 and transport costing \$720. Both records are valid, but the total cost of \$3,600 exceeds the budget. A repair that keeps the selected transport leaves \$2,280 for family accommodation. None of the available accommodation options meets both requirements, so selecting again from the same options will not resolve the problem. Additional local research is needed to find suitable accommodation before itinerary assembly can continue.

This additional research can reuse the route and transport selection. Previously collected accommodation records also remain available, although they do not provide a suitable option for this repair. The new options must meet the accommodation requirements, and the revised itinerary must pass the budget and final-output checks. The distinction between these two cases is whether the existing inputs are sufficient for repair. This determines whether to retry the current action or add work before continuing.

\section{\tool: Contribution-Guided Repair Framework}
\subsection{Overview}
\label{sec:method-overview}

Building on the example in Section~\ref{sec:illustrative-example}, we propose \tool, a contribution-guided runtime repair framework for MAS applications. Figure~\ref{fig:overview} shows how \tool uses unmet task requirements and accepted results to select repairs and continue execution.

The MAS Specification Extractor describes the methods available in the host MAS, including their inputs, outputs, and conditions of use. The Task Modeler combines this MAS specification with the user request to build a contribution graph and configure verification policies. It uses an existing plan when one is available. Each subtask has an expected contribution: the result it should produce and the acceptance conditions that result must meet. The graph connects these results to the subtasks that use them. Framework adapters connect \tool to the host MAS so that it can observe actions and apply repairs during execution.

\tool checks actions and results against the method requirements and verification policies. When an action fails, action-level repair attempts to produce the required result using the existing inputs. Contribution assessment then uses the checking records to determine what remains to be done. Results that pass their checks can support later subtasks. If an input is still being produced by a planned subtask, \tool schedules or waits for that subtask. If repair requires a missing prerequisite or a change to unfinished work, graph-level repair adds or revises subtasks and their dependencies.

Both repair levels reuse results that still satisfy their acceptance conditions. When available, scenario repair templates guide the choice of changes and parameters. \tool checks the repair outputs, reassesses task progress, and continues execution. Once the task requirements are met, it generates and verifies the final answer.

\begin{figure}[tbp]
    \centering
    \includegraphics[width=0.9\linewidth]{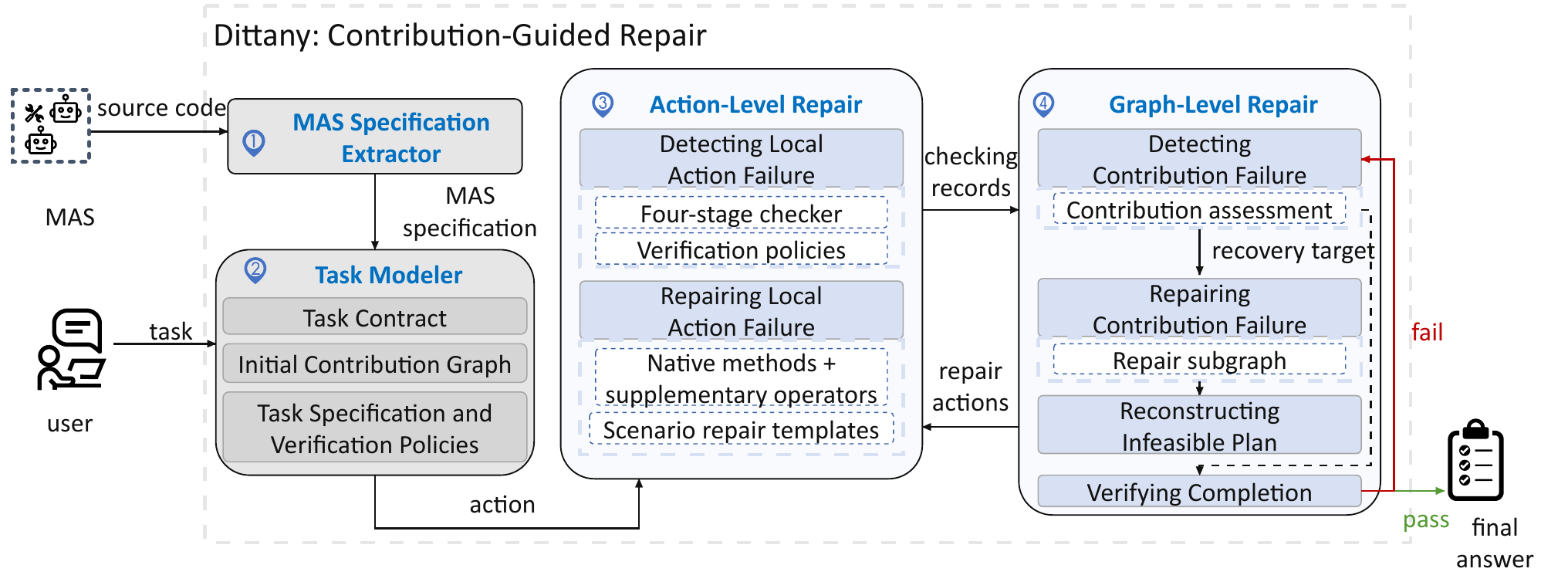}

    \caption{Overview of \tool.}
    \label{fig:overview}
\end{figure}

\subsection{MAS Specification Extractor}

\tool first constructs a specification describing the methods available in the host MAS and their conditions of use. The Task Modeler uses these descriptions to connect task requirements to executable work. Runtime checking and repair use the same descriptions to assess invocations and returned results, select methods, and connect their inputs. Specifically, the \emph{MAS specification} records (1) agent definitions and instructions; (2) the executable operations available to them, termed \emph{methods}; (3) each method's callable reference, input parameters, defaults and types, output structure, behavior, input and output meanings, and conditions of use; and  (4) output-to-input dependencies in any declared workflow. A method is exposed through the application's interface for an agent invocation or tool invocation. \tool retains references to the supporting code or text with the extracted information.

\emph{Agent definitions and instructions.} \tool statically scans the source code to locate agent constructor calls and their configuration. \tool follows the configuration references to record each agent's definition and associated instructions.

\emph{Methods, interfaces, behavior, and conditions of use.} \tool follows configuration references, data dependencies, and the call graph to find the implementations of each agent's methods. \tool extracts callable references and parameter names, defaults, and types from function signatures, and output fields from declared schemas and explicitly constructed return objects. \tool uses an LLM to describe each method's behavior and the meaning of its inputs and outputs from the implementation, agent instructions, and documentation. \tool extracts conditions of use from explicit checks and stated restrictions, keeping references to the code or text that supports each description.

\emph{Workflow dependencies.} If the application declares a workflow, \tool follows references from one node's output to another node's input to record the declared dependencies for the Task Modeler. \tool combines these dependencies with the agent definitions, instructions, available methods, and method descriptions to form the MAS specification.

After obtaining the specification, \tool validates it through an LLM that compares the generated descriptions with the corresponding code, instructions, and documentation, identifying contradictions and unsupported claims. \tool revises the flagged descriptions before using the specification in planning, action checking, and repair.

\medskip
\emph{An illustrative example.} In Figure~\ref{fig:travel-specification}, \tool identifies the concierge and its search and calculation tools from the source code. It describes \texttt{plan\_trip} as consuming a route skeleton and grounded local information to produce a complete trip plan. The Task Modeler uses these descriptions to connect the producers of those inputs to itinerary generation.

\begin{figure}[tbp]
    \centering
    \begin{minipage}[t]{0.65\linewidth}
        \vspace{0pt}
        \centering
        \includegraphics[width=\linewidth]{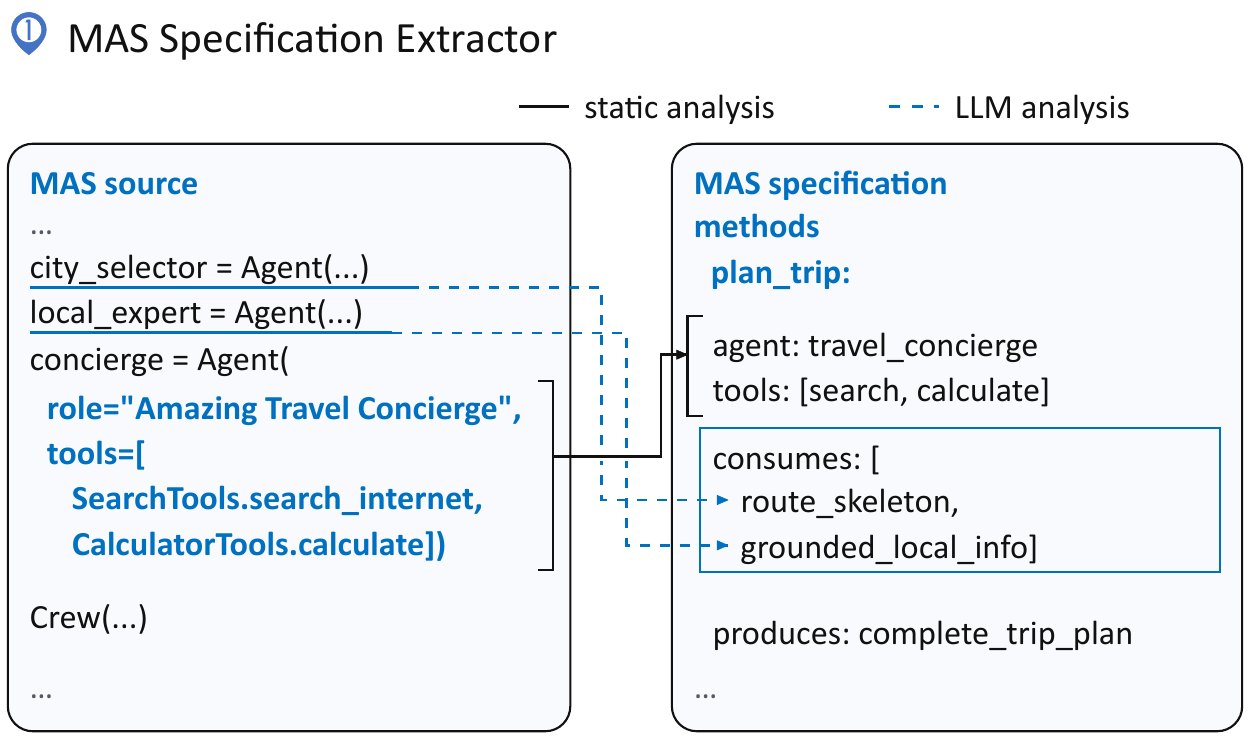}
        \caption{MAS specification extracted from the travel-planning application's source code.}
        \label{fig:travel-specification}
    \end{minipage}\hfill
    \begin{minipage}[t]{0.335\linewidth}
        \vspace{0pt}
        \centering
        \includegraphics[width=\linewidth]{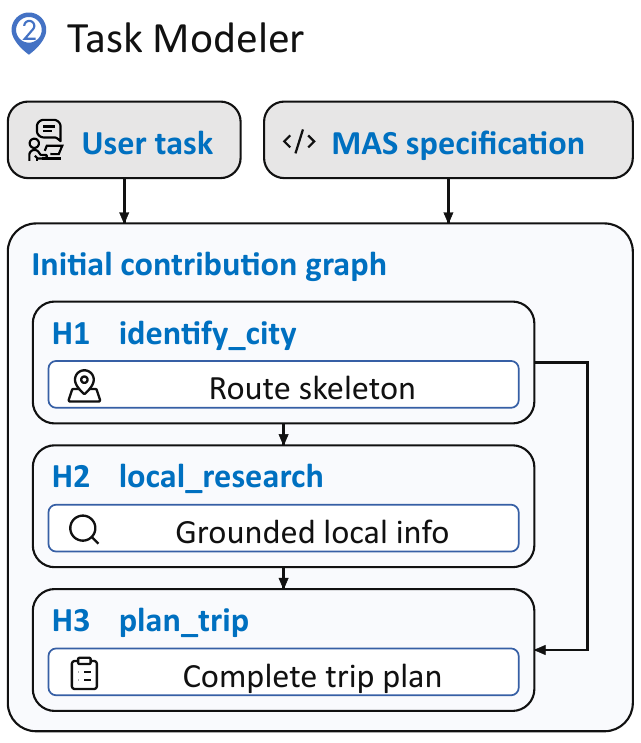}
        \caption{Initial contribution graph for the travel-planning task.}
        \label{fig:travel-task-model}
    \end{minipage}
\end{figure}

\subsection{Task Modeler}
\label{sec:task-modeler}

The MAS specification describes what the application can execute; the user request determines what the current task must achieve. \tool combines the MAS specification with the user request to model the task as a contribution graph. It extracts the task requirements, identifies the results and methods that can satisfy them, connects result dependencies, and configures verification policies. These policies guide runtime checking, while the graph relates checking records to task progress and the work needed for repair.

\subsubsection{Generating Task Contract}

\tool uses an LLM to extract the requested deliverable, content, and restrictions, retaining their source instructions. It distinguishes explicit obligations and limits from explicitly stated preferences, then compares the extraction with the request using an LLM to correct omissions or changed meanings. \tool records the reviewed requirements as the \emph{task contract}: mandatory requirements define completion, while preferences guide choices among suitable plans and repairs.

\subsubsection{Constructing Initial Contribution Graph}

Using the task contract, \tool identifies the work needed to produce each required result. For an existing plan, it links steps to method descriptions in the MAS specification and uses an LLM to match their outputs to task requirements by information, object, and scope. Uncovered requirements identify results needing additional steps. When no plan exists, \tool starts from all required results. For each result lacking a producing step, it creates a subtask and selects a method with a matching output and conditions of use suited to the task context.

\tool then matches each selected method's required inputs to request values, available results, or planned outputs using the same semantic comparison. It binds available values directly and connects planned outputs to their receiving inputs. For an input without a suitable source, \tool adds a producing subtask and repeats method and input selection. It records these subtasks and connections as a \emph{contribution graph}: nodes specify expected results, and edges connect outputs to receiving inputs. \tool programmatically validates method and parameter references, source types, and dependency order, returns incompatible choices for revision, and records unresolved choices. The expected results remain subject to runtime verification.

\subsubsection{Generating Task Specification and Verification Policies}

The graph identifies how required results can be produced. To determine whether these results can support subsequent work, \tool also establishes their acceptance conditions. \tool derives \emph{acceptance conditions} for these results from their linked mandatory task requirements and receiving methods' input requirements. It uses an LLM to express the conditions and check consistency with those sources, retaining each source requirement and any associated consuming method and input. \tool attaches each condition to the result or combination it concerns. A subtask's expected result and acceptance conditions define its \emph{expected contribution}; the required results and conditions, including those for the final deliverable, form the \emph{task specification}.

\tool turns each acceptance condition into executable checks. Using an LLM and supplied procedure descriptions, it identifies the required judgment and supporting inputs, then selects procedures whose operations and input types express that judgment: schema checkers compare fields and types, numerical evaluators compute expressions and test bounds, and reference checkers resolve records and compare fields. For judgments requiring semantic interpretation beyond these operations, \tool configures an LLM-based checker with the condition and supporting material. It combines procedures when one condition requires several judgments.

\tool then binds these procedures' operands and supporting material to request fields, available results and planned outputs by meaning and scope. When data or evidence lacks a source, it selects a producing method, adds a subtask, connects its output to the check, and derives conditions and checks for the new result. \tool programmatically validates references, types, supported operations, and configurations, returning invalid bindings or procedures for revision. It records each condition, configured procedures, and bindings as a \emph{verification policy}, linked to the relevant results and graph nodes. During execution, these links identify the policies applicable to each action and returned result. The resulting checking records support action-level repair and contribution assessment, connecting an unmet condition to the relevant result and its downstream uses.

\medskip
\emph{An illustrative example.} In Figure~\ref{fig:travel-task-model}, the subtasks cover city selection ($H_1$), local research ($H_2$), and itinerary generation ($H_3$). The route from $H_1$ supplies $H_2$ and $H_3$, while the accommodation and transport options from $H_2$ supply $H_3$. From the travel requirements and method inputs, \tool derives a completeness condition for $H_2$'s accommodation and transport information. For $H_3$, it configures a numerical evaluator to sum the selected costs and compare the total with the \$3,000 limit.

\subsection{Action-Level Repair}
\label{sec:action-repair}

\tool treats one execution of a selected method within a subtask as an \emph{action}, such as an agent invocation or tool invocation. Using the MAS specification and the Task Modeler's verification policies, it checks the action against its assigned result and applicable requirements. The failed conditions and available inputs guide action-level repair when the required result can be obtained within the existing dependencies. Checking records also enter contribution assessment (Section~\ref{sec:graph-repair}) to determine how the result affects the remaining task.

\subsubsection{Detecting Local Action Failure}
\label{sec:detect-action}

\tool applies a four-stage checker at method selection, input preparation, execution, and return. Before dispatch, it associates the action with a subtask through execution context or the mapping procedure in Section~\ref{sec:graph-repair}. From the action's arguments or delegation instructions and the method's output description, \tool identifies its assigned result. It resolves the agent and method in the MAS specification and compares the method's behavior and conditions of use with that result and the task requirements, using an LLM for semantic judgments.

\tool then follows the subtask's input bindings to compare the required sources with the supplied arguments and context. It checks source availability and acceptance, validates required parameters, types, and explicit bounds, and compares references with the bound objects, versions, and scopes. These checks are programmatic, including reference resolution in the receiving agent's context; \tool uses an LLM to compare content with the method's input descriptions and subtask requirements. Missing source information defers the action and enters contribution assessment (Section~\ref{sec:graph-repair}); available information omitted from the action or inaccessible to the receiving agent calls for input or handover repair.

Once these checks pass, \tool allows the action to execute and inspects its status and reported errors. For a returned result, it checks fields and types against the output interface and follows the Task Modeler's result-to-policy links to check the assigned content, including omissions and relations with earlier inputs used by the action. \tool binds actual values and evidence to the procedures and executes them; semantic checks return an LLM judgment with supporting or conflicting passages and missing evidence. Requirements assigned to other actions and independent result combinations enter contribution assessment.

\tool records the affected action, argument, or result field with the checked condition, values, and evidence. It labels a condition satisfied when all required checks pass, violated when a check demonstrates failure, or unestablished when information is insufficient; pending checks resume when their inputs arrive. \tool marks content as \emph{grounded} when checks establish its required support and source consistency. It retains results or parts satisfying their applicable conditions as \emph{accepted results}, with evidence and check outcomes. The failed conditions and accepted results guide candidate construction below. Contribution assessment uses the same records to determine what the action has established, what remains unmet, and whether further inputs or work are needed.

\subsubsection{Repairing Local Action Failure}
\label{sec:repair-action}

The checking records identify the unmet condition and the values or content involved. \tool uses these records to construct candidate repairs that address the condition with the available inputs. For method-selection failures, it proposes method substitution or an agent with access to the required method. For invalid or incomplete arguments, it proposes value adjustments or supplying omitted available content; for inconsistent or inaccessible references, it proposes corrections or input rebinding to suitable available sources. For execution errors, it considers retry when the error and method conditions support repetition, alongside action adjustment or method substitution. For output-interface or content failures, it selects methods to correct or regenerate the affected result. Conditions awaiting planned inputs remain pending for contribution assessment.

For method substitution or content correction, \tool uses an LLM to match alternative methods' outputs to the required result and their conditions of use to the task context, then selects an agent with access to the method. Alongside native methods, \tool provides three executable methods with task-specific parameters, termed \emph{supplementary operators}:

\noindent\textit{Recomputation.} \tool retrieves the expression, operand bindings, and source values identified by the failed calculation check, evaluates the expression with a supported evaluator, and corrects the affected field with the computed value.

\noindent\textit{Evidence-constrained revision.} \tool supplies an LLM with the affected content, its unmet acceptance condition, and relevant evidence to correct claims, add supported content, or reorganize the result. The revised content undergoes the same verification policy.

\noindent\textit{Constrained reassignment.} \tool identifies adjustable choices, candidate values, and fixed selections from the task contract and failed check. It uses an LLM to propose assignments, checks their mandatory requirements with the configured procedures, and ranks passing assignments by task preferences. Numerical bounds and explicit relations are evaluated programmatically; new selections retain references to unchanged source records.

\tool uses a common configuration process for candidates at both repair levels, drawing on their method descriptions, input requirements, and available results. The application may supply \emph{scenario repair templates}: reusable configurations of applicability conditions, preferred changes, and parameter guidance. \tool matches templates to the failure and task state, then uses an LLM with matching templates, the candidate, failed condition, available data, and repair history to propose parameters and input bindings. Inputs reference request values, accepted results, or affected content and evidence. The original policies supply expressions and acceptance conditions; the task contract supplies fixed requirements and preferences. These inputs determine the result that the configured repair must establish.

\tool checks each configuration against the MAS specification and available data, validating references, types, bounds, and source suitability, with an LLM for semantic comparisons. When a method or input use changes, it replaces conditions derived solely from the old use, retains mandatory task requirements, and reconfigures affected policies through the Task Modeler before checking reused results. Invalid configurations return for revision; inputs needing further production enter contribution assessment. \tool excludes previously failed choices with unchanged inputs and conditions.

\tool orders eligible candidates using an LLM to compare how they address the failure, guided by template preferences and previous outcomes. It executes the first candidate within budget and reapplies the action checks and verification policies. Unresolved conditions and new evidence guide another candidate; exhaustion of eligible candidates or budget ends action-level repair. Each checked return or execution update, including deferral and repair termination, updates contribution assessment (Section~\ref{sec:graph-repair}).

When action-level repair ends with unmet conditions, \tool reassesses the remaining producers and dependencies. It schedules or waits for viable planned producers, and invokes graph-level repair only when recovery requires missing prerequisites or changes to unfinished work and sufficient recovery budget remains. If local candidates are exhausted and no such change is identified, \tool records the recovery target as unresolved. Ending action-level repair does not reset the applicable budgets or make previously failed candidates eligible again under unchanged inputs and conditions.

\medskip
\emph{An illustrative example.} In Figure~\ref{fig:travel-action-repair}, action $a_1$ in $H_2$ returns a truncated structured response. \tool detects the incomplete output and retries \texttt{local\_research} as action $a_2$ using $H_1$'s accepted route. It checks the returned accommodation and transport options before passing them to $H_3$.

\begin{figure}[tbp]
    \centering
    \includegraphics[width=\linewidth]{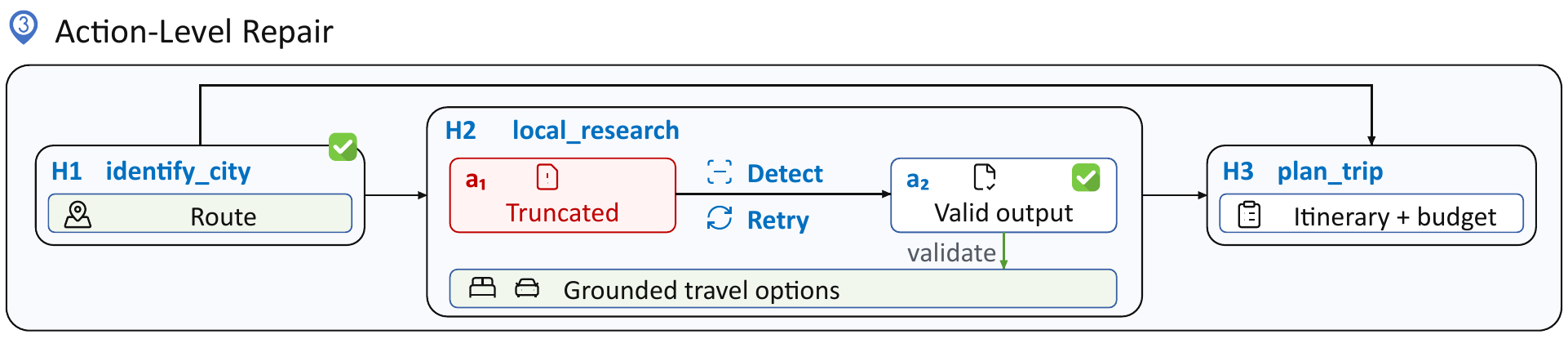}
    \caption{Action-level repair of a truncated local-research response.}
    \label{fig:travel-action-repair}
    \medskip
    \includegraphics[width=\linewidth]{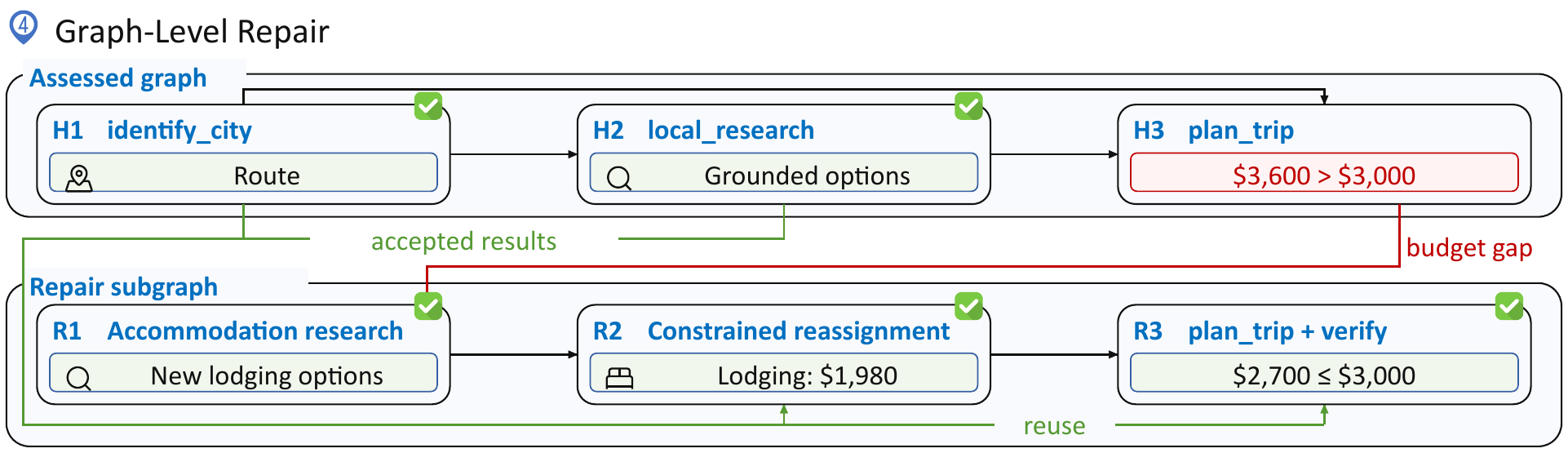}
    \caption{Graph-level repair of the budget violation through accommodation research and reassignment.}
    \label{fig:travel-graph-repair}
\end{figure}

\subsection{Graph-Level Repair}
\label{sec:graph-repair}

Action-level checking establishes what an invocation has produced. Determining whether the task can proceed also requires relating that result to the requirements and dependencies of the remaining work. \tool uses \emph{contribution assessment} to relate action-level checking records to the remaining task through the contribution graph. After each checked return or execution update, it compares accumulated results with the subtasks' expected contributions and assesses the inputs available to unfinished work. This assessment determines whether execution can continue with accepted results, proceed after planned inputs arrive, or needs graph-level repair to add subtasks or revise unfinished work. For graph-level repair, the unmet conditions and reusable results define the recovery target described below.

\subsubsection{Detecting Contribution Failure}

\tool associates actions with subtasks through execution context or the host's node mapping. Otherwise, it uses an LLM to compare the action's assigned result and expected output with active nodes' expected results and identify a unique match. After each checked return or execution update, \tool adds the results, checking records, and execution status to that subtask.

If no unique association can be established, \tool retains the action's observations and checking records without crediting them to a graph node. It uses available calling context, input sources, and expected outputs to refine the association within the semantic-call budget. An unresolved association neither closes a node nor releases work dependent on that result; independent work can continue. A result passing its own checks remains distinct from an established contribution to a particular node.

\tool reuses these records to identify grounded content and satisfied conditions, and executes pending checks on independently produced result combinations when their inputs arrive. It compares accumulated content with the expected result and consumer input requirements by information, object, and scope, following result references and using an LLM for semantic matching. Content supplying the required information is \emph{useful} and becomes available after its applicable checks pass. Missing content and unestablished or violated conditions form the \emph{contribution gap}. Individually accepted results retain their evidence and checking records when a combination fails; revised candidates and selections are recorded separately. The unmet combination condition guides the changes needed, while the accepted results remain candidates for reuse under their applicable conditions.

An unmet condition may still be resolved by work already in progress. \tool therefore assesses progress before deciding whether additional work is needed. \tool compares results and checking records before and after each update for progress: additional accepted results or usable evidence for unmet requirements, newly satisfied conditions or corrected violations, and verified exclusions advancing a choice. It assesses each checked return or execution update once; a monitoring window without feedback from an executing subtask counts as one assessment without progress. Three consecutive assessments without progress constitute stagnation; progress resets the count.

Using the assessed results and progress, \tool closes a node when its results provide the expected result and satisfy all acceptance conditions, then schedules successors with available inputs. Incomplete subtasks continue while making progress. For missing inputs, \tool follows bindings to producing subtasks, schedules unstarted producers with ready inputs, or waits while they progress. During this wait, it pauses the receiving subtask's stagnation count and monitors the producer; a stalled or failed producer enters the assessment below.

When a subtask with unmet conditions ends, stagnates, or cannot continue, \tool examines whether the remaining plan can satisfy those conditions. It compares the required results with unfinished nodes' expected outputs, input bindings, and execution status. Viable producers follow the scheduling and waiting decisions above. A missing producer or a necessary change to a subtask or dependency triggers graph-level repair. \tool records the unmet conditions and their policies, reusable results and evidence, and affected unfinished nodes as the \emph{recovery target}, specifying what repair needs to establish and where its outputs connect.

\subsubsection{Repairing Contribution Failure}
\label{sec:repair-contribution}

\tool uses the recovery target from contribution assessment to determine the results needed for repair, drawing on its unmet conditions, checked values, and evidence. It uses an LLM to compare what each condition requires with the accepted results: missing content calls for its production, unestablished support calls for the evidence required by the policy, and a demonstrated violation identifies the content or selection to revise. \tool retains these links between required results and unmet conditions for checking the proposed repair. The required results define the work to establish, and accepted results provide the inputs already available. \tool supplies these results, the MAS specification, unfinished plan, and repair history to an LLM to propose recovery methods.

\tool selects and configures proposed native methods and supplementary operators through the common process in Section~\ref{sec:repair-action}. Their output descriptions must match the required information, objects, and scope, and their conditions of use must fit the task context. Matching scenario repair templates guide parameter choices and candidate preferences.

\tool traces each selected method's required inputs using their meaning, type, and conditions of use in the MAS specification. It uses an LLM to match them to request values, accepted results, or outputs assigned to planned subtasks. \tool binds available values directly and connects inputs requiring planned outputs to their producing nodes. For an input with no suitable source, it selects a method whose output can supply the required information, adds a producing subtask, and repeats input selection for that method's prerequisites. The resulting subtasks and dependencies form a \emph{repair subgraph}. \tool derives acceptance conditions and configures verification policies for new results as in Section~\ref{sec:task-modeler}, and records which recovery outputs supply unfinished nodes.

If no applicable method can supply a required input, \tool rejects that candidate repair and tries alternative methods or configurations. During recursive input construction, it tracks the required results and conditions on the current construction path. A candidate that depends on a result still being constructed on that path is rejected as cyclic; a suitable planned producer is reused rather than duplicated. The same failure handling applies when constructing sources for verification inputs. \tool excludes candidates already unsuccessful under unchanged inputs and conditions. If alternatives are exhausted or the applicable budget is reached, it retains accepted results and reports the unmet conditions without supplying unsupported values or relaxing mandatory requirements.

\tool first checks whether the proposed subgraph covers the recovery target. For each unmet condition, it locates the proposed result and the inputs required by its verification policy. \tool uses an LLM to compare the proposed output with the required information, objects, and scope, and programmatically validates condition references and verification-input sources. If a required result is uncovered, \tool returns that requirement to method selection; if verification data or evidence lacks a source, it extends input construction to obtain it.

\tool checks the configured methods and bindings through the common configuration checks in Section~\ref{sec:repair-action}, and validates dependency order across the repair subgraph. It revises incompatible choices and adds producing subtasks for missing prerequisites. Configured checks with available inputs further identify choices that violate their conditions. Once these checks pass, the repair subgraph provides the proposed results and input sources needed to address the recovery target.

\subsubsection{Reconstructing Infeasible Plan}

To resume execution with these repair outputs, \tool determines which nodes in the unfinished plan can continue with the proposed inputs and which need to change. It starts with each unfinished node that will receive a recovery output. It compares the proposed input with the node's method requirements and conditions of use through the same interface and semantic checks. When the method can produce its expected result using that input, \tool retains the node and updates its binding. For missing input information, it extends the repair subgraph to supply it. When the method is unsuitable, \tool adds the node's expected result and unmet conditions to the recovery target and selects replacement subtasks.

\tool follows output dependencies from each changed node and repeats these checks for affected unfinished successors. It updates conditions and policies for changed input uses as described in Section~\ref{sec:repair-action} and checks reused results accordingly. \tool retains independent nodes and accepted results whose conditions remain satisfied, together with their source records. It incorporates affected subtasks into the repair subgraph and binds retained results to the revised work. If rechecking invalidates a previously accepted result, \tool no longer counts the affected contribution as complete and adds its required result and unmet conditions to the recovery target for regeneration. The same assessment applies to completed consumers whose results depended on the invalidated result. Independent contributions whose conditions remain satisfied are retained.

\tool schedules the resulting subtasks as their inputs become available within the recovery budget. The same action checks and action-level repair apply to their execution. Newly checked results update contribution assessment, which determines whether subsequent work can proceed or further repair is needed. If recovery cannot continue within budget, \tool reports unresolved conditions with retained results.

If a checked repair leaves conditions unmet, \tool updates the recovery target using the new results and evidence, removing conditions already satisfied. Further repair requires a viable candidate under the updated state and remaining budget. When no eligible candidate remains, \tool reports the unresolved conditions with the accepted results instead of repeating an unchanged failed repair.

\subsubsection{Verifying Completion}

\tool follows the links from task-contract requirements to checking records to assess completion. Once mandatory conditions checkable before final-answer generation are satisfied, it uses a complete host response or initiates final-answer generation with accepted results, even if the host has not submitted. \tool evaluates the actual answer, including an earlier host submission, using the final-deliverable verification policies to check required content and structure, consistency with accepted results, and conditions affected by generation. It applies action-level repair to defects confined to the current action, and returns missing underlying results or required changes to unfinished subtasks to contribution assessment for graph-level repair. \tool checks the revised deliverable before returning it.

\emph{An illustrative example.} In Figure~\ref{fig:travel-graph-repair}, $H_3$ combines supported lodging and transport records costing \$2,880 and \$720, but the \$3,600 itinerary violates the \$3,000 budget. \tool reuses the accepted route and transport, leaving a lodging budget of \$2,280. The existing accommodation options do not meet both this limit and the family-accommodation requirements, so \tool adds accommodation research ($R_1$). Constrained reassignment ($R_2$) uses the new options and retained candidates to select lodging costing \$1,980. $R_3$ then combines the retained results and selected lodging into a \$2,700 itinerary and verifies it against the task requirements.

\section{Implementation}
We implement \tool as a Python runtime library for applications built with AutoGen/AG2, CrewAI, LangGraph, Google ADK, HelloAgents, and the OpenAI Agents SDK. Framework integrations expose the host's available methods, execution results, and control operations to a common checking and repair process. The contribution graph and verification policies operate on this common view, allowing the same recovery procedure to be used across the six frameworks.

To use \tool, developers provide the application source, execution entry point, and runtime configuration. \tool derives the MAS specification and builds the task model for each user request. Applications can supply domain verification procedures for requirements such as numerical consistency, itinerary feasibility, and source-supported claims. Scenario repair templates express applicability conditions, candidate preferences, and parameter guidance for recurring tasks. These procedures and templates connect domain requirements to the common verification and recovery process.

The library connects runtime checking to method selection, input preparation, execution, and return. Checks run before results supply subsequent work, and their outcomes update contribution assessment. The same execution facilities invoke native methods and supplementary operators, then check their outputs and connect them to the remaining task. Recovery retains validated results under the conditions of their subsequent use and charges additional work to the configured budget. Extending \tool to another framework requires exposing that framework's methods, observations, and supported control operations; extending it to a new domain uses verification procedures and scenario templates.

\section{Evaluation}
We evaluate whether \tool can recognize unmet task requirements, recover correct task outcomes, and benefit from combining two repair scopes. Our evaluation addresses three questions:
\begin{itemize}
    \item \textbf{RQ1 (Detection):} How accurately does \tool detect failures in MAS execution?
    \item \textbf{RQ2 (Repair):} How effectively does \tool improve task correctness compared with Native execution and recovery baselines?
    \item \textbf{RQ3 (Ablation):} How do graph-level and action-level repair contribute to task accuracy and token cost?
\end{itemize}

\subsection{Methodology}
\label{sec:eval-methodology}

\subsubsection{Applications and Test Data}
We evaluate 30 open-source MAS applications, with five from each of AutoGen/AG2, CrewAI, LangGraph, Google ADK, HelloAgents, and the OpenAI Agents SDK. We collect applications from GitHub and framework example repositories as of July 1, 2026, selecting those with runnable multi-agent entry points, assessable outcomes, and more than 100 stars in their hosting repositories. They span research and content generation (8), software engineering (3), travel planning (3), customer and benefits support (3), financial analysis (6), education (1), operations (2), and data retrieval and decision support (4).

Table~\ref{tab:eval-repair-apps} lists the applications, which total 25,460 LOC and 151 agent roles. Application sizes range from 59 to 6,147 LOC (median 430), with 2--21 roles (median 4). LOC counts nonblank, noncomment application Python lines, excluding docstrings, tests, third-party libraries, and deployment code. Roles count distinct reasoning agents, with repeated invocations counted once and deterministic orchestration and nonreasoning user proxies excluded.

We prepare 300 inputs and reference outcomes per application, yielding 9,000 application--task instances. Research tasks use World Bank indicators~\cite{worldbankindicators}, Nobel Prize records~\cite{nobelprizedata}, Crossref~\cite{crossrefrestapi}, QASPER~\cite{dasigi2021qasper}, and OpenAlex~\cite{priem2022openalex}. Software tasks use SWE-bench-Live~\cite{zhang2025swebenchlive} and MBPP-sanitized~\cite{austin2021mbpp}. Trip Planner and Travel Concierge adapt TravelPlanner~\cite{xie2024travelplanner}; AG2 Travel Planner uses requests over a fixed place catalog and route snapshot with feasible itineraries.

Customer Service adapts 50 tau2-bench~\cite{barres2025tau2} airline tasks into six environment variants; Healthcare Support combines 50 Synthea~\cite{walonoski2018synthea} patient records with six support templates. Financial tasks use FinQA~\cite{chen2021finqa}, frozen yFinance market data, and financial disclosures. Education and operations use MathDial~\cite{macina2023mathdial}, QASC~\cite{khot2020qasc}, NetOpsBench~\cite{yang2026netopsbench}, and RCAEval~\cite{pham2024rcaeval}. The remaining tasks use Wine Quality~\cite{cortez2009winequality}, Chinook~\cite{chinookdatabase}, resumes~\cite{resumedataset}, job descriptions~\cite{drushchak2024djinni}, and FEVER~\cite{thorne2018fever}, balanced across its three labels. Inputs preserve the application task and use frozen source records or initial environments. Reference outcomes are reserved for offline evaluation.

\subsubsection{Baselines}
We compare \tool with the host application without additional recovery (\emph{Native}) and four recovery techniques:
\begin{enumerate}
    \item \emph{SupervisorAgent}~\cite{lin2025supervisoragent} uses an LLM-free filter to trigger supervisory review and corrective guidance.
    \item \emph{AgentTether}~\cite{zhao2026agenttether} uses dependency-aware trajectory diagnosis and repair memory to guide re-execution.
    \item \emph{Reflexion}~\cite{shinn2023reflexion} uses verbal feedback to improve subsequent attempts. We adapt it to the MAS level by providing the task, trajectory, and application-visible feedback, and adding the reflection to the next attempt's context.
    \item \emph{Iterative Rollback (IR)} restores the latest supported checkpoint preceding a failure and re-executes the remaining work, moving to an earlier checkpoint after persistent failure until success or budget exhaustion.
\end{enumerate}
We compare complete systems with their respective checking and recovery mechanisms on shared task inputs and source snapshots. On the five applications marked in Table~\ref{tab:eval-repair-apps}, SupervisorAgent and AgentTether allow up to three attempts, including the initial execution. SupervisorAgent selects from predefined interventions; AgentTether uses LLM-based diagnosis with repair memory disabled. Their other 25 results use the reported configurations and assessments.

\subsubsection{Assessment and Metrics}
We assess detection against the task requirements and evidence available at the corresponding interaction, assigning one label per execution. Detection precision, recall, F1, and accuracy use 8,758 resolved labels; 164 ambiguous runs and 78 runs with format-only discrepancies are excluded. A true positive requires identifying an actual failure at the corresponding interaction, independently of its category or repair outcome.

Task accuracy is the fraction of inputs whose final outputs satisfy the reference requirements. Reported percentages are rounded to two decimal places; accuracy gains are calculated from task counts, and relative token changes from token totals, before rounding. Numerical relations, program tests, artifacts, and environment changes use executable checks; semantic assessment uses DeepSeek V4 Flash with thinking disabled, the task, candidate output, rubric, and frozen evidence. The same reference-outcome assessment applies to Native, \tool, Reflexion, and IR, and to SupervisorAgent and AgentTether on the five marked applications. Paired outcomes compare Native and \tool executions of the same input. To examine requirements associated with these outcomes, we group Native failures using their reference checks: an unmet volatility requirement in Stock Analysis, missing inter-activity routes in AG2 Travel Planner, and correct claim values with incomplete source bindings in Agents Deep Research. Each group includes every task satisfying its definition; subsequent success requires passing the complete task assessment.

We measure deployment cost in tokens consumed by application execution, checking, and recovery, excluding offline assessment. For executions without complete usage measurements, we estimate tokens from observed call counts and the costs of comparable calls or applications, applying the same comparable-call rule across methods. We report recorded and estimated tokens over all 9,000 inputs, using the existing estimates for missing usage.

\subsubsection{Execution Settings}
Experiments use an Intel Core i5-14600KF processor (14 cores, 20 threads), 32 GB RAM, a 1 TB GIGABYTE G325E1TB drive, Windows 10 Pro (build 19045), and Python 3.12.3. Runtime inference uses DeepSeek V4 Flash through remote APIs, with temperature 0 and thinking disabled.

\subsection{Answer to RQ1: Detection}
\label{sec:eval-detection}

Across 8,758 executions with resolved detection labels, \tool reports 4,257 failures, of which 3,758 are true positives and 499 are false positives. It detects 3,758 of the 4,746 executions containing failures, achieving 88.28\% precision and 79.18\% recall (Table~\ref{tab:eval-detection}). These metrics assess whether an actual failure is detected at the relevant interaction; successful repair is evaluated separately in RQ2.

\begin{table*}[t]
\centering
\begin{minipage}[t]{\textwidth}
\vspace{0pt}
\caption{Run-level failure detection across six frameworks. Precision, recall, F1, and accuracy are percentages over resolved labels under the evaluation protocol.}
\label{tab:eval-detection}
\centering
\footnotesize
\setlength{\tabcolsep}{3pt}
\begin{tabular}{@{}lrrrr@{}}
\hline
Framework & Precision & Recall & F1 & Accuracy \\
\hline
AutoGen/AG2 & 83.27 & 72.12 & 77.29 & 64.69 \\
CrewAI & 65.52 & 90.66 & 76.07 & 81.58 \\
Google ADK & 95.91 & 89.18 & 92.42 & 96.67 \\
HelloAgents & 100.00 & 65.40 & 79.08 & 70.15 \\
LangGraph & 95.06 & 96.53 & 95.79 & 95.60 \\
OpenAI Agents & 92.57 & 84.51 & 88.36 & 88.88 \\
\hline
Overall & 88.28 & 79.18 & 83.48 & 83.02 \\
\hline
\end{tabular}
\end{minipage}\hfill
\end{table*}

\paragraph{Detection across failure types and applications.}
We distinguish single-point defects, including invalid tool calls and malformed outputs, from defects in task content, including incorrect calculations, unsupported claims, and violated planning constraints. Among the 3,758 true-positive executions, single-point defects are documented in 1,534 (40.82\%) across 21 applications, while task-content defects are documented in 1,064 (28.31\%) across 14 applications. Both types span all six frameworks. These counts record documented occurrences throughout execution, including recovery, and are nonexclusive: 71 executions contain both types.

The task-content detections establish a common need across applications: checking whether an agent has returned an output also requires checking what that output contributes to the task. Numerical claims must agree with their source observations, claims must have the requested evidence, and planned activities must satisfy their joint constraints. The detections across 14 applications show that these requirements extend beyond the validity of individual invocations and output formats.

For illustration, Code Generator rejects a malformed planner response before the developer consumes it, whereas Stock Analysis identifies a missing required metric in a report that declares itself complete. The latter execution already has the prices needed for the calculation; the unmet task condition identifies what remains to be corrected. These cases illustrate the two checking scopes observed across applications.

\noindent\emph{Finding.} \tool achieves 88.28\% precision and 79.18\% recall. Its detections include task-content defects across 14 applications and all six frameworks, alongside single-point defects across 21 applications.

\subsection{Answer to RQ2: Repair}
\label{sec:eval-repair}

\paragraph{Overall effectiveness.}
\tool achieves 84.76\% task accuracy, compared with 58.01\% for Native and 63.48\% for Reflexion, the strongest aggregate recovery baseline (Table~\ref{tab:eval-repair}). Across 9,000 tasks, the gains over Native and Reflexion are 26.74 and 21.28 percentage points, corresponding to 2,407 and 1,915 additional correct outcomes, respectively.

\begin{table*}[t]
\centering
\footnotesize
\setlength{\tabcolsep}{5pt}
\caption{Task accuracy (\%) across six frameworks, with five applications and 1,500 tasks per method in each framework. Bold marks the highest accuracy among all six methods. IR denotes Iterative Rollback; Superv. denotes SupervisorAgent; Tether denotes AgentTether. Application-level results appear in Table~\ref{tab:eval-repair-apps}.}
\label{tab:eval-repair}
\begin{tabular}{lrrrrrr}
\hline
Framework & Native & Reflexion & IR & Superv. & Tether & \tool \\
\hline
AutoGen/AG2 & 52.60 & 75.33 & 56.53 & 60.13 & 59.53 & \textbf{93.67} \\
CrewAI & 40.60 & 27.27 & 28.00 & 19.67 & 22.73 & \textbf{93.07} \\
Google ADK & 82.93 & 73.93 & 73.67 & 34.80 & 49.13 & \textbf{88.67} \\
HelloAgents & 60.87 & 63.60 & 55.80 & 48.67 & 49.40 & \textbf{86.13} \\
LangGraph & 64.13 & \textbf{86.20} & 80.33 & 49.13 & 48.40 & 77.33 \\
OpenAI Agents & 46.93 & 54.53 & 54.93 & \textbf{77.93} & 73.27 & 69.67 \\
\hline
Overall & 58.01 & 63.48 & 58.21 & 48.39 & 50.41 & \textbf{84.76} \\
\hline
\end{tabular}
\end{table*}

In the paired executions, \tool passes 2,838 of the 3,779 inputs that Native fails (75.10\%), while Native passes 431 that \tool fails. Both pass 4,790 inputs and both fail 941.

\paragraph{Breadth and additional correct outcomes.}
The improvement extends to 24 of the 30 applications and all six frameworks; two applications tie with Native and four have lower accuracy (Table~\ref{tab:eval-repair-apps}). Omitting any single application leaves an aggregate gain of at least 24.22 percentage points over Native. \tool leads the recovery techniques in four frameworks; Reflexion leads on LangGraph, and SupervisorAgent on OpenAI Agents.

The gains also include outcomes missed by both reflection and rollback: 1,342 of the 2,838 inputs that \tool passes and Native fails (47.29\%) also fail under Reflexion and IR. These additional correct outcomes span 23 applications and all six frameworks.

\paragraph{Recovery guided by unmet contributions.}
The gains cover numerical consistency, connections between activities, and source-supported claims. Among 279 Stock Analysis tasks with an unmet Native volatility requirement, \tool passes 276, compared with 4 for Reflexion and 5 for IR. It passes all 114 AG2 Travel Planner tasks with missing routes between activities, compared with 65 and 33, respectively. In Agents Deep Research, it passes 79 of 86 tasks with correct Native claim values but incomplete source bindings, compared with 53 and 59. Each group includes all tasks meeting its definition; success requires passing the complete task assessment. These results show that recovery addresses missing task contributions even when individual outputs are already present or claim values are correct.

The trajectories show how these unmet conditions guide recovery. In the EQX analysis, \tool computes the missing volatility metric from already observed prices: 30 daily log returns yield 59.84\% annualized volatility. Recovery updates that metric while preserving the other analyses and recommendation. The corresponding Reflexion and IR outputs give 55.6\% and 62.0\%, respectively, and fail the same numerical requirement.

Missing connections require additional work. In the Milan itinerary, \tool retains the selected venues and creates a recovery stage to supply their absent walking connection. The resulting 34-minute, 2.8-km leg completes an itinerary satisfying all ten constraints, compared with nine for Native.

Missing source support similarly directs a two-stage research recovery: acquire the evidence, then revise and check the report. The three existing section drafts remain inputs, and their writers are not re-executed. The completed report provides the required claim-to-source binding and passes the task assessment. Across these executions, the unmet requirement supplies recovery's goal, while completed contributions supply inputs to the additional work.

\begin{table*}[t]
\centering
\scriptsize
\setlength{\tabcolsep}{3pt}
\caption{Application size, agent roles, and task accuracy (\%) for all 30 applications, with 300 tasks per application and method. LOC and agent counts follow the definitions in the setup. The Overall row sums LOC and roles across applications. Bold marks the best of all six methods, including ties. $\ast$: SupervisorAgent and AgentTether use the three-attempt configurations described in the setup; their other 25 rows use reported baseline results.}
\label{tab:eval-repair-apps}
\label{tab:application-inventory}
\begin{tabular}{llrrrrrrrr}
\hline
Framework & Application & LOC & Agents & Native & Reflexion & IR & Superv. & Tether & \tool \\
\hline
AutoGen/AG2 & E-commerce & 263 & 5 & 29.67 & 32.67 & 28.00 & 0.00 & 0.00 & \textbf{100.00} \\
 & Travel Planner$^{\ast}$ & 437 & 4 & 33.00 & 63.00 & 27.67 & 97.67 & 90.33 & \textbf{100.00} \\
 & AutoGen AGI & 1,275 & 10 & 93.33 & \textbf{99.67} & \textbf{99.67} & 96.67 & 99.33 & 95.33 \\
 & Financial Analysis & 357 & 4 & 99.67 & 98.00 & 98.33 & 99.33 & \textbf{100.00} & \textbf{100.00} \\
 & Research Agents 3.0 & 139 & 3 & 7.33 & \textbf{83.33} & 29.00 & 7.00 & 8.00 & 73.00 \\
\hline
CrewAI & Devyan & 198 & 4 & 70.00 & 0.00 & 0.00 & 32.67 & 33.67 & \textbf{70.33} \\
 & Easy Investment & 1,135 & 4 & 99.67 & 90.33 & 92.67 & 40.67 & 48.00 & \textbf{100.00} \\
 & Profile Match$^{\ast}$ & 59 & 2 & 20.33 & 30.67 & 34.67 & 25.00 & 32.00 & \textbf{98.33} \\
 & Stock Analysis & 318 & 3 & 2.67 & 1.33 & 2.00 & 0.00 & 0.00 & \textbf{98.67} \\
 & Trip Planner & 321 & 4 & 10.33 & 14.00 & 10.67 & 0.00 & 0.00 & \textbf{98.00} \\
\hline
Google ADK & Academic Research & 184 & 3 & 92.00 & 98.00 & \textbf{98.67} & 65.33 & 66.00 & 92.33 \\
 & LLM Auditor & 184 & 2 & 41.00 & \textbf{62.67} & 62.00 & 47.67 & 46.33 & 62.33 \\
 & Software Bug Assistant & 124 & 2 & 98.00 & \textbf{100.00} & \textbf{100.00} & 27.00 & 33.33 & 97.67 \\
 & Financial Advisor$^{\ast}$ & 475 & 5 & \textbf{96.67} & 82.00 & 80.67 & 2.33 & 48.00 & 91.00 \\
 & Travel Concierge & 1,496 & 21 & 87.00 & 27.00 & 27.00 & 31.67 & 52.00 & \textbf{100.00} \\
\hline
HelloAgents & DeepCast & 2,152 & 4 & 0.00 & 27.00 & 26.33 & 0.00 & 0.00 & \textbf{100.00} \\
 & LearningAgent$^{\ast}$ & 1,693 & 7 & 93.00 & 85.33 & 52.00 & 54.00 & 51.33 & \textbf{93.67} \\
 & Network Health Report$^{\ast}$ & 1,030 & 5 & \textbf{100.00} & 64.33 & 58.67 & 82.33 & 89.00 & \textbf{100.00} \\
 & SRE On-Call & 573 & 3 & 14.00 & 42.67 & \textbf{44.00} & 7.00 & 6.67 & 40.33 \\
 & StockSage & 6,147 & 5 & 97.33 & 98.67 & 98.00 & \textbf{100.00} & \textbf{100.00} & 96.67 \\
\hline
LangGraph & Code Generator & 372 & 3 & 70.00 & \textbf{86.67} & 66.00 & 59.33 & 56.67 & 82.33 \\
 & Deep Agent & 62 & 2 & 71.00 & 65.67 & 69.67 & 66.33 & 62.00 & \textbf{88.00} \\
 & Music Store & 254 & 3 & 83.33 & \textbf{89.67} & 75.67 & 77.67 & 82.00 & 85.67 \\
 & Open Deep Research & 1,513 & 6 & 50.33 & 94.33 & \textbf{96.00} & 4.67 & 7.00 & 65.67 \\
 & Researcher & 751 & 6 & 46.00 & \textbf{94.67} & 94.33 & 37.67 & 34.33 & 65.00 \\
\hline
OpenAI Agents & Agents Deep Research & 1,528 & 10 & 0.00 & 50.33 & 55.67 & \textbf{85.00} & 56.67 & 74.00 \\
 & Customer Service & 719 & 8 & 65.00 & 67.00 & \textbf{69.67} & 61.33 & 61.00 & 68.00 \\
 & Financial Research & 423 & 6 & 23.00 & 68.67 & 70.33 & \textbf{78.33} & \textbf{78.33} & 61.00 \\
 & Healthcare Support & 1,057 & 4 & 79.00 & 56.67 & 57.33 & 83.33 & \textbf{89.00} & 77.67 \\
 & Research Bot & 221 & 3 & 67.67 & 30.00 & 21.67 & \textbf{81.67} & 81.33 & 67.67 \\
\hline
Overall & 30 applications & 25,460 & 151 & 58.01 & 63.48 & 58.21 & 48.39 & 50.41 & \textbf{84.76} \\
\hline
\end{tabular}
\end{table*}

\noindent\emph{Finding.} \tool improves task correctness across 24 applications, including 1,342 tasks on which Native, Reflexion, and IR all fail. The examined trajectories show requirement-guided recovery that retains usable analysis and supplies missing computations, connections, or source support.

\subsection{Answer to RQ3: Ablation}
\label{sec:eval-ablation-scope}

We compare Full with Action-only and Graph-only configurations (Table~\ref{tab:eval-ablation}). Action-level repair corrects the current action within its existing dependencies; graph-level repair revises the remaining work and its dependencies. Full supports both scopes. Native and Full share their task accuracy and deployment costs with the main comparison.

\begin{table}[t]
\centering
\small
\caption{Task accuracy and estimated deployment tokens across repair configurations. Costs include application execution, checking, and recovery; offline assessment is excluded.}
\label{tab:eval-ablation}
\begin{tabular}{lrrr}
\hline
Configuration & Passed tasks & Success (\%) & Tokens (M) \\
\hline
Native & 5,221 & 58.01 & 317.963 \\
Action-only & 4,729 & 52.54 & 325.932 \\
Graph-only & 6,225 & 69.17 & 451.397 \\
Full & \textbf{7,628} & \textbf{84.76} & 374.948 \\
\hline
\end{tabular}
\end{table}

\paragraph{Graph-level repair increases task correctness.}
Adding graph-level repair to Action-only raises accuracy from 52.54\% to 84.76\%, producing 2,899 additional correct task outcomes, or a gain of 32.21 percentage points. Graph-level repair expands recovery to work that must be added or revised to satisfy the task. The accuracy gain demonstrates the value of this broader scope alongside correction of the current action.

\paragraph{Action-level repair reduces token consumption.}
Adding action-level repair to Graph-only reduces estimated deployment tokens from 451.397M to 374.948M, saving 76.449M tokens (16.94\%). Accuracy simultaneously increases from 69.17\% to 84.76\%. Action-level repair offers focused correction within existing dependencies before recovery expands to the task graph. Its inclusion therefore improves the measured correctness--cost tradeoff: the combined configuration completes 1,403 more tasks with fewer tokens than Graph-only.

\noindent\emph{Finding.} Graph-level repair increases accuracy by 32.21 percentage points over Action-only; action-level repair reduces token consumption by 16.94\% relative to Graph-only while further improving accuracy. Together, they achieve 84.76\% task accuracy.

\subsection{Deployment Cost}
\label{sec:eval-cost}

Table~\ref{tab:eval-deployment-cost} compares task accuracy and deployment tokens across all six methods over the full 9,000-task suite. \tool uses 374.948M tokens, 17.92\% more than Native, while producing 2,407 additional correct outcomes. Among the recovery techniques, it achieves the highest aggregate accuracy with the lowest token consumption: it uses 24.19\% fewer tokens than Reflexion, 13.05\% fewer than IR, 24.20\% fewer than SupervisorAgent, and 16.93\% fewer than AgentTether. Thus, the correctness gains are accompanied by lower total token cost than every recovery baseline in this comparison.

\begin{table}[t]
\centering
\small
\caption{Task accuracy and estimated deployment tokens over all 9,000 tasks. Tokens include application execution, checking, and recovery.}
\label{tab:eval-deployment-cost}
\begin{tabular}{lrr}
\hline
Method & Accuracy (\%) & Tokens (M) \\
\hline
Native & 58.01 & 317.963 \\
Reflexion & 63.48 & 494.565 \\
IR & 58.21 & 431.231 \\
SupervisorAgent & 48.39 & 494.658 \\
AgentTether & 50.41 & 451.353 \\
\tool & \textbf{84.76} & 374.948 \\
\hline
\end{tabular}
\end{table}

\section{Threats to Validity}
\noindent\textbf{Specification and checking.}
\tool derives method descriptions and task conditions from source code, documentation, and user requests. Incomplete descriptions or ambiguous requests can lead to omitted requirements or incorrect associations. Semantic checks also depend on the model and supplied evidence. These errors can affect which results are accepted and whether repair remains within existing dependencies or adds work. Reuse therefore depends on the coverage of the configured checks and whether their conditions still apply to the new use of a result.

\noindent\textbf{Outcome assessment.}
Executable checks and model-assisted judgments cover different aspects of task completion. Detection assessment relates failure labels to task requirements and the evidence at the corresponding interactions. Semantic judgments can contain correlated errors or admit multiple interpretations, and excluding unresolved labels limits detection conclusions to the resolved subset.

\noindent\textbf{Execution and configuration.}
Model responses, external tools, and service updates can change execution trajectories and repair opportunities. Temperature zero does not ensure identical responses across repeated remote calls. Checking templates, confidence thresholds, and repair budgets influence which conditions are observed and which repairs can be attempted. The main comparison evaluates complete systems with their respective checking and recovery mechanisms; the repair-level ablation fixes \tool's checking configuration to compare repair configurations.

\noindent\textbf{Application coverage and observability.}
The evaluated applications and six frameworks cover a finite set of tasks and execution patterns. Their selection limits generalization to other MAS applications. Repair also requires sufficient observations of actions, results, and dependencies. Unobservable state or external effects can prevent the verification needed to reuse a result or continue affected work.

\noindent\textbf{Cost measurement.}
We evaluate deployment token consumption over all 9,000 tasks, combining recorded usage with estimates based on observed calls where complete token measurements are unavailable. Native and \tool use the same estimation rule. These totals capture application execution, checking, and repair; monetary cost and elapsed time additionally depend on service pricing, latency, and concurrent execution.

\section{Related Work}
\subsection{Failure Analysis and Attribution in Agent Systems}
Cemri et al.~\cite{cemri2025why} study MAS failures involving specification and system design, inter-agent alignment, and task verification and termination. Zhang et al.~\cite{zhang2025whowhen} formulate automated failure attribution as identifying the responsible agent and decisive step, and introduce the Who\&When dataset. DoVer examines ambiguity in failure attribution and tests hypotheses through targeted interventions and re-execution~\cite{ma2025dover}. AgentTether constructs a dependency-aware graph of observed transitions, identifies failure-critical subtrajectories, and uses the diagnosis to guide recovery~\cite{zhao2026agenttether}.

\tool establishes a recovery target by comparing required results and acceptance conditions with accumulated results and remaining producers. Planned work that can still supply a result leads to scheduling or waiting; a missing prerequisite or an unsuitable continuation calls for graph-level repair. The target links each unmet condition to the results needed to address it, the accepted inputs available, and the unfinished consumers that must receive the repair outputs.

\subsection{Runtime Fixing in Agent Systems}
Reflexion converts execution feedback into verbal reflections retained to guide subsequent attempts~\cite{shinn2023reflexion}. For code generation, Self-Debugging uses execution and natural-language explanations to revise generated programs~\cite{chen2024selfdebug}. SupervisorAgent introduces adaptive runtime supervision, using an LLM-free filter to trigger corrective guidance and observation revision~\cite{lin2025supervisoragent}. These techniques use feedback to improve generated artifacts or ongoing behavior.

Other approaches explicitly repair execution trajectories. DoVer modifies messages or plans at selected points and resumes execution to assess the intervention~\cite{ma2025dover}. MARS combines rollback, guided repair, and continuation in a Monte Carlo Tree Search over candidate repairs~\cite{lu2026mars}. AgentTether combines diagnosis and repair memory with optional guarded runtime intervention during re-execution~\cite{zhao2026agenttether}. In \tool, the available inputs and unmet conditions determine whether repair can remain within the current action. When further prerequisites are needed, graph-level repair constructs their producers and checks how the new outputs connect to unfinished consumers. Reuse is conditional on the retained results satisfying the requirements of their subsequent use; a changed use triggers reassessment before execution continues.

Software integration research also addresses failures in AI-enabled execution. Keeper tests software that uses ML APIs and suggests changes to API usage~\cite{wan2025keeper}. SmartGear detects and converts incompatible ML API outputs using expectations derived from surrounding software and consistency across related invocations~\cite{wan2023smartgear}. Comfrey adapts LLM and RAG responses to software requirements concerning format, syntax, and repetition~\cite{shao2026comfrey}. These works establish the importance of the software context in checking and correcting AI outputs. In MAS applications, satisfying a task condition can also require producing missing information and revising dependencies among unfinished subtasks. \tool connects action-level correction to this recovery work, checking both repair outputs and their use in subsequent execution.

\section{Conclusion}
We presented \tool, a runtime repair framework for MAS applications. It uses unmet task requirements, accepted results, and their dependencies to determine the work needed to recover a partially failed execution. Action-level repair obtains the required result within existing dependencies, while graph-level repair adds or revises work when those dependencies need to change. Accepted results supply recovery and subsequent execution under their applicable conditions. Each checked result updates the assessment, guiding the application from a partially successful execution toward a verified final answer. Across 30 MAS applications and 9,000 task instances, \tool achieves 84.76\% task accuracy, compared with 58.01\% for Native and 63.48\% for Reflexion, the strongest aggregate recovery baseline. Across all 9,000 tasks, estimated deployment tokens increase from 317.963M for Native to 374.948M for \tool, an increase of 17.92\%. The repair-level ablation supports combining action-level correction with graph-level recovery to improve task success and token efficiency.

\section*{Acknowledgment}
This paper is supported by National Natural Science Foundation of China (Grant No. 62402183, 92582108) and Fundamental Research Funds for the Central Universities.

\bibliographystyle{ACM-Reference-Format}
\bibliography{citation.bib}

\end{document}